\documentclass[a4,oneside]{amsart}

\usepackage{graphicx, verbatim,amsmath,amssymb,array,setspace}
\usepackage[colorlinks=true, pdfstartview=FitV, linkcolor=blue, citecolor=green, urlcolor=black,filecolor=magenta]{hyperref}

\newcommand{\ppp}{{\mathcal P}}

\newcommand{\rrr}{{\mathcal R}}

\newcommand{\T}{{\bf T}}
\newcommand{\R}{{\mathbb R}}

\newcommand{\N}{{\mathbb N}}

\newcommand{\vecx}{{{\bf x}}}

\def\blueb{{B}}
\def\reda{{A}}
\def\greent{{T}}

\newtheorem{thm}{Theorem}[]
\newtheorem*{thm*}{Theorem}

\newtheorem{dfn}[thm]{Definition}

\everymath{\displaystyle}

\theoremstyle{definition}
\newtheorem{ex}{Example}
\newtheorem*{ex*}{Example}

\begin{document}

\title[Fusion tilings]{Fusion: a general framework for hierarchical tilings}
\author[Natalie Priebe Frank] % (optional, use only with lots of authors)
{Natalie Priebe Frank}
\address{Natalie Priebe Frank\\Department of Mathematics\\Vassar
  College\\Poughkeepsie, NY 12604} \email{nafrank@vassar.edu}
%\thanks{Thank you all very much...} 
%\date{June 29, 2012}
%\date{ ICQ12, September 1, 2013}

\begin{abstract}
One well studied way to construct quasicrystalline tilings is via inflate-and-subdivide (a.k.a. substitution) rules. These produce self-similar tilings--the 
Penrose, octagonal, and pinwheel tilings are famous examples. We present a different 
model for generating hierarchical tilings we call ``fusion rules". Inflate-and-subdivide 
rules are a special case of fusion rules, but general fusion rules are more flexible and allow for defects, changes in geometry, and even constrained randomness.  A condition that produces homogeneous structures and a method for computing frequency for fusion tiling spaces are discussed.

%allow the composition 
%rules to change from level to level. Spectral, dynamical, and topological results will be 
%discussed that parallel those known for self-similar tilings.

\end{abstract}

\maketitle

\section{Introductory examples}
Suppose you are making a polymer out of building blocks $A$ and $B$, perhaps they are atoms or proteins or molecules or some other objects, but for our purposes they will be called tiles.  Suppose further that these building blocks assemble into longer and longer blocks in stages, according to the following rules.  At the first stage, they assemble into two-tile patches of the form $AB$ and $BA$.  These are called `1-supertiles' and can be thought of as unit cells.  The set of 1-supertiles is denoted $\ppp_1 = \{AB, BA\}$.  In the second stage, $1$-supertiles assemble into strings of the form $AB\,BA$ and $BA\,AB$, which can be thought of as supercells.  We denote the set of 2-supertiles as $\ppp_2 = \{ABBA, BAAB\}$. At the third stage the 2-supertiles assemble to form the strings $ABBA\,BAAB$ and $BAAB\,ABBA$, giving us $\ppp_3 = \{ABBABAAB,BAABABBA\}$.

This process continues indefinitely by building up new supertiles as concatenations of supertiles from the previous level.  For this example, if we denote the first $n$-supertile to be $\ppp_n(1)$ and the second to be $\ppp_n(2)$, then the set of $(n+1)$-supertiles takes the form $\ppp_{n+1} = \{\ppp_n(1)\ppp_n(2), \ppp_n(2)\ppp_n(1)\}$.    Note that the fact that $AA$ is not a 1-supertile does not preclude it from appearing in a 2-supertile.  Moreover, note that the fusion rules given here are in no way intrinsic; rather they are choices---different rules can use different choices.

We wish to consider all infinite strings of $A$s and $B$s that are made up of arbitrarily large supertiles.  An infinite string (tiling) $\T$ is said to be {\em admitted by} the fusion rule if every finite substring appearing in $\T$ can be found in either $\ppp_n(1)$ or $\ppp_n(2)$ if $n$ is taken to be sufficiently large.  This particular fusion rule generates the so-called Thue-Morse or Prouhet-Thue-Morse sequences, which look like
$$\cdots BABAABABBAABBABAABBAABABBAABBABAABABB\cdots$$

In the terminology of this paper we call the $AB\,BA$ a {\em fusion} of $AB$ and $BA$, which in one dimension is the same as concatenation but will be more general in higher dimensions.  A {\em fusion rule} is the set of all possible supertiles: $\rrr = \{\ppp_n \text{ such that } n \in \N\}$.  The fusion rule contains all possible patches and so governs the allowed structure.

\begin{figure}[ht]
\includegraphics[width=.2in]{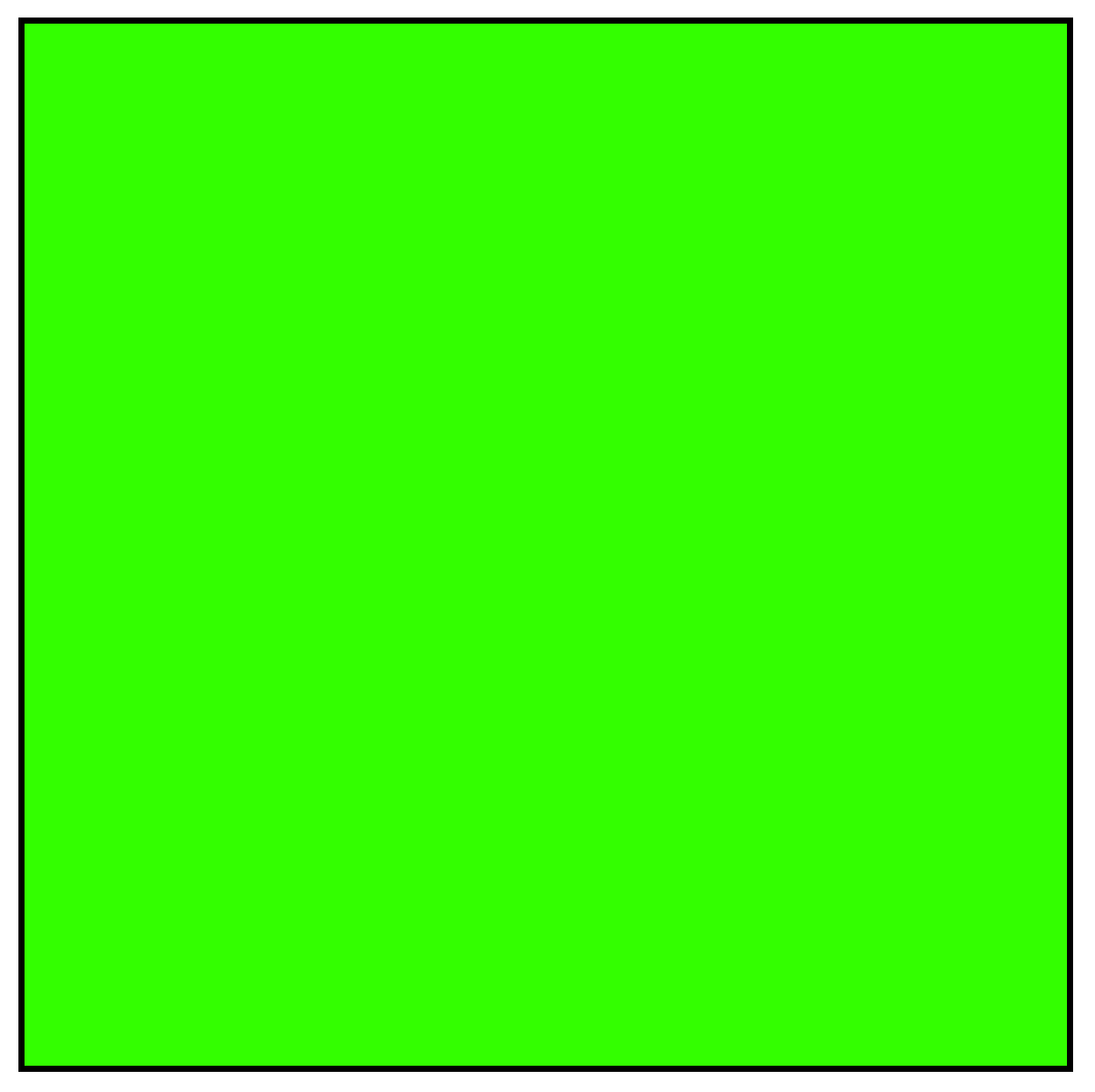}\hfill \includegraphics[width=.2in]{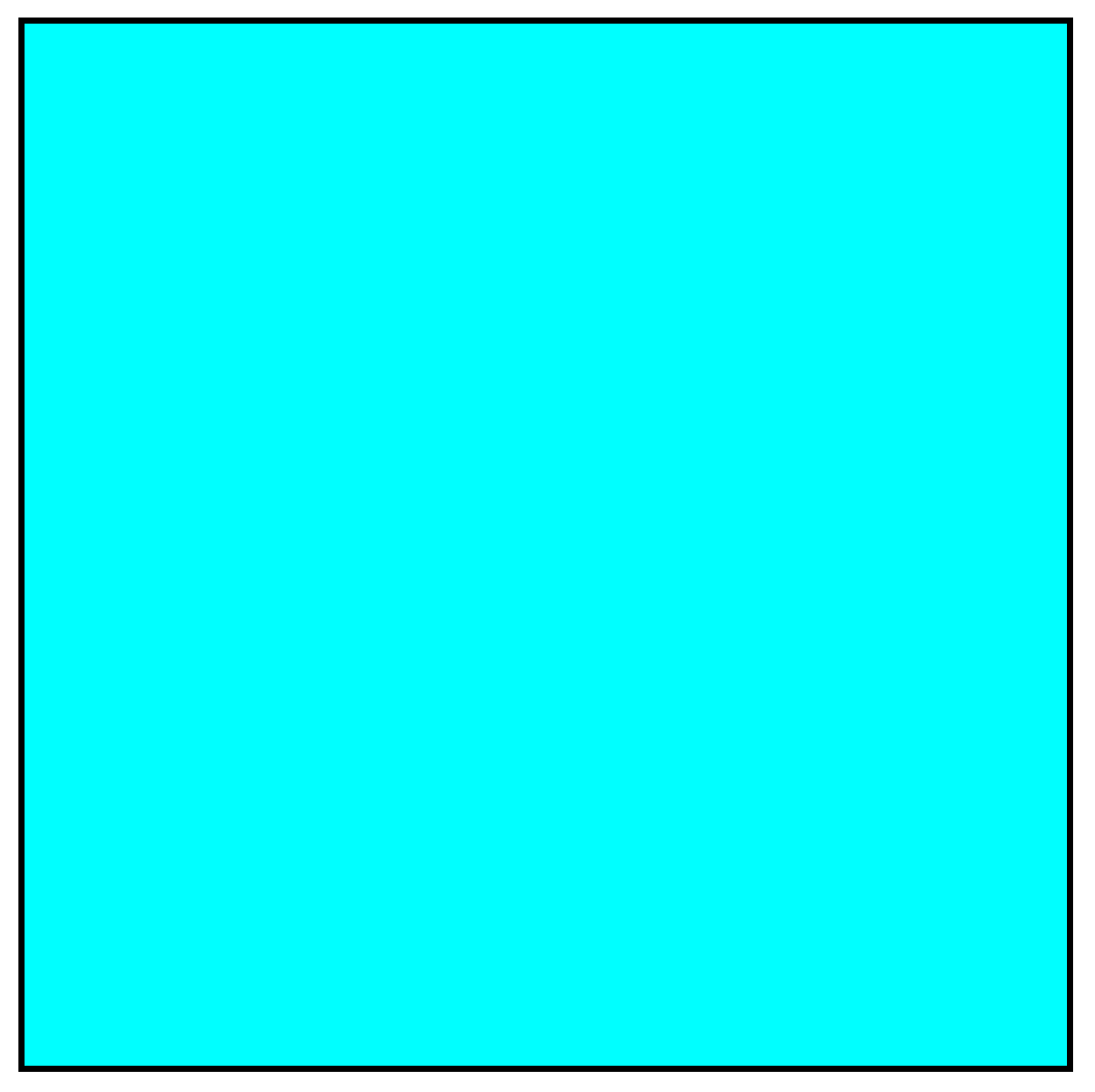}\hfill \includegraphics[width=.2in]{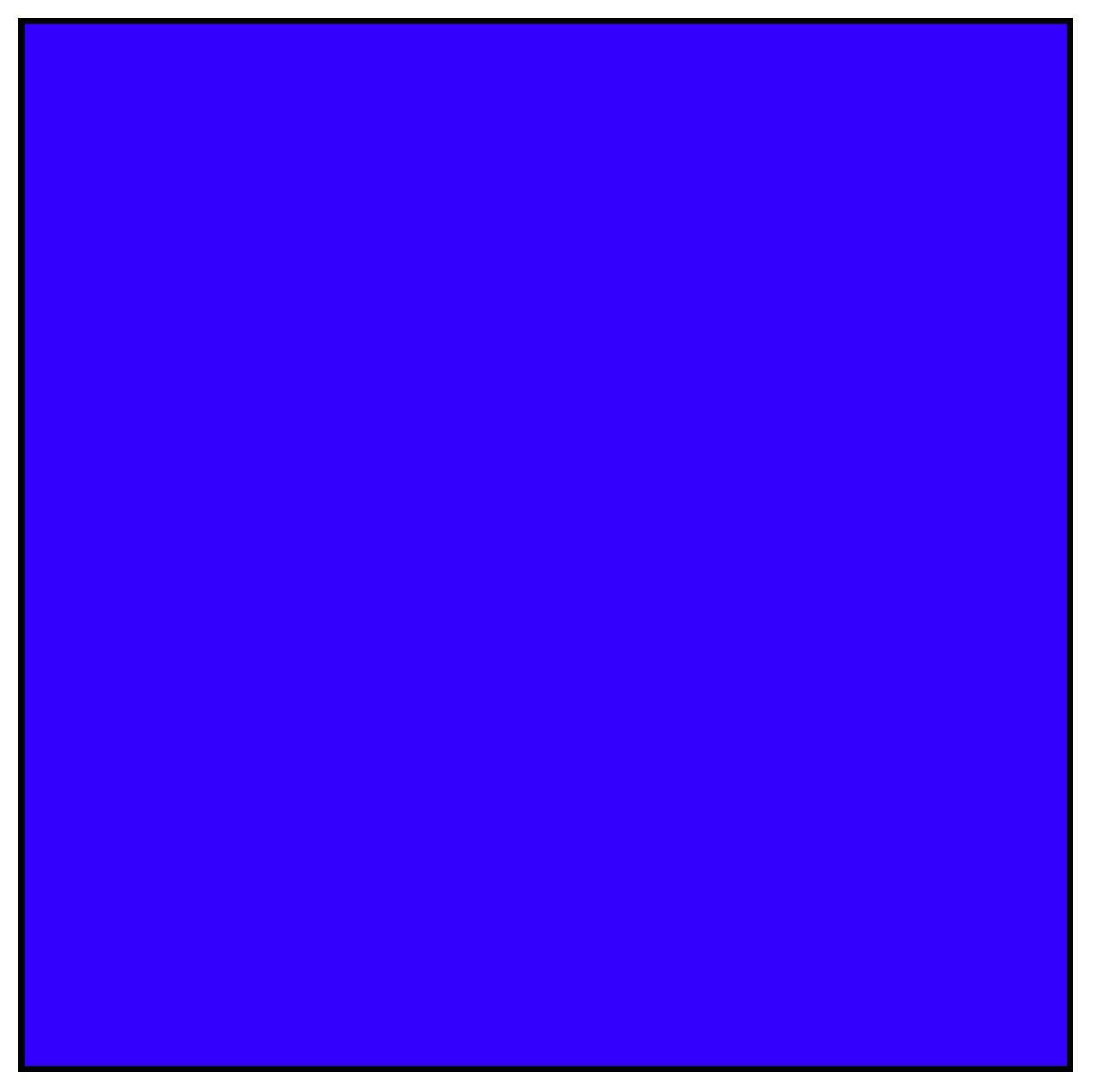}\hfill \includegraphics[width=.2in]{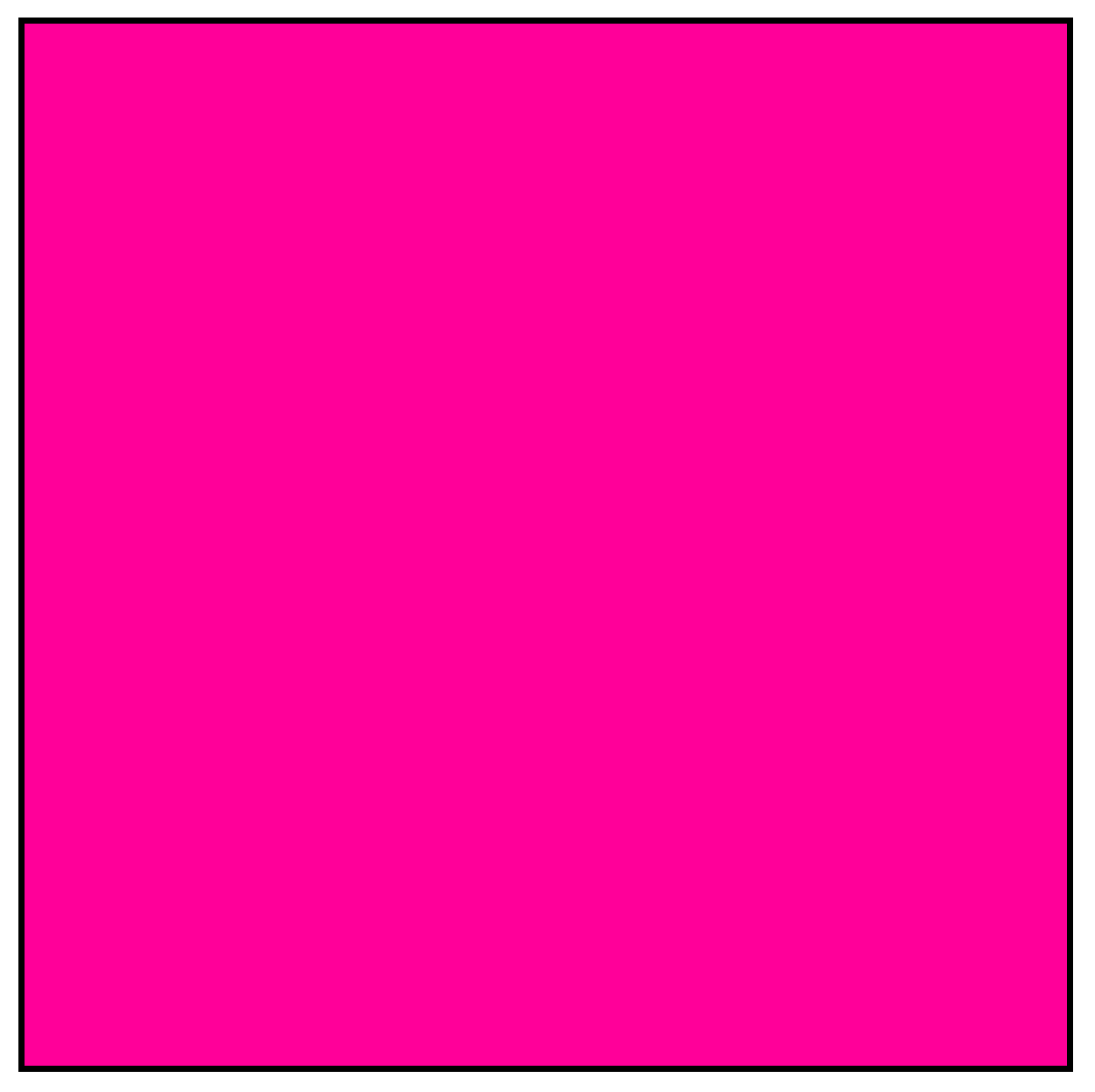}\hfill
\caption{The four prototile types.}
\label{fibtiles}
\end{figure}

For a two-dimensional example, consider a tiling made from unit squares of four different types, denoted in figure \ref{fibtiles} in color (greyscale) and thought of as atoms.  We give a sample fusion rule for your consideration, but you might enjoy tinkering with the rules to make different tilings.

\begin{figure}[ht]
\includegraphics[width=.4in]{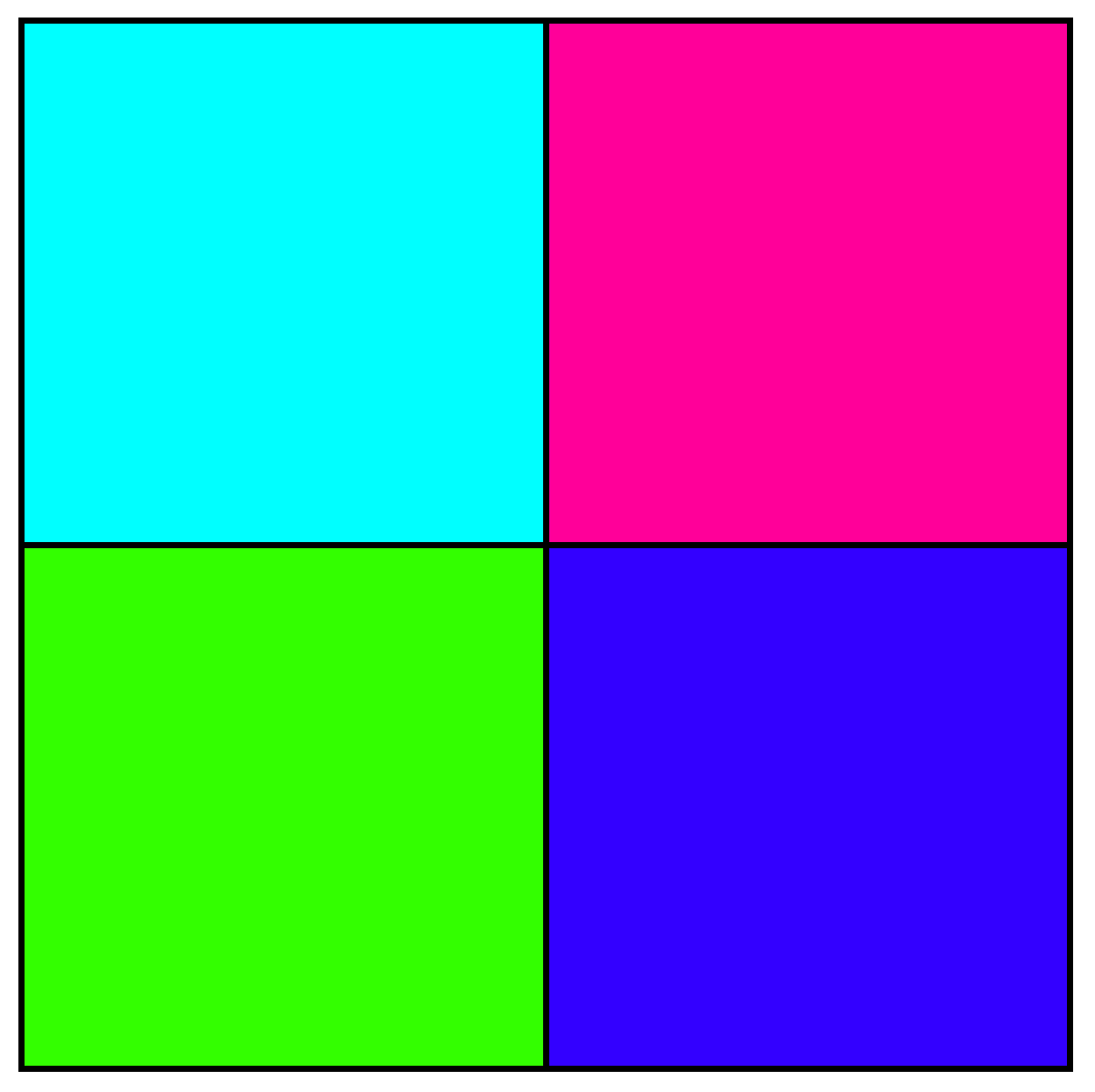}\hfill \includegraphics[width=.4in]{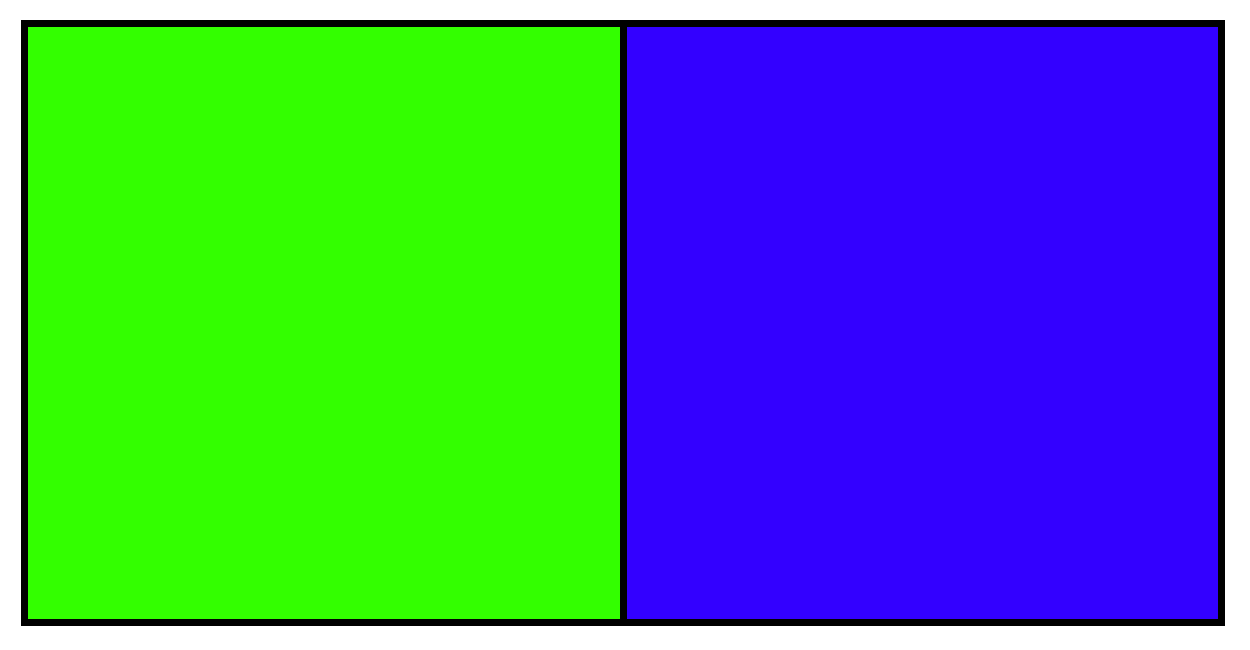}\hfill 
\includegraphics[width=.2in]{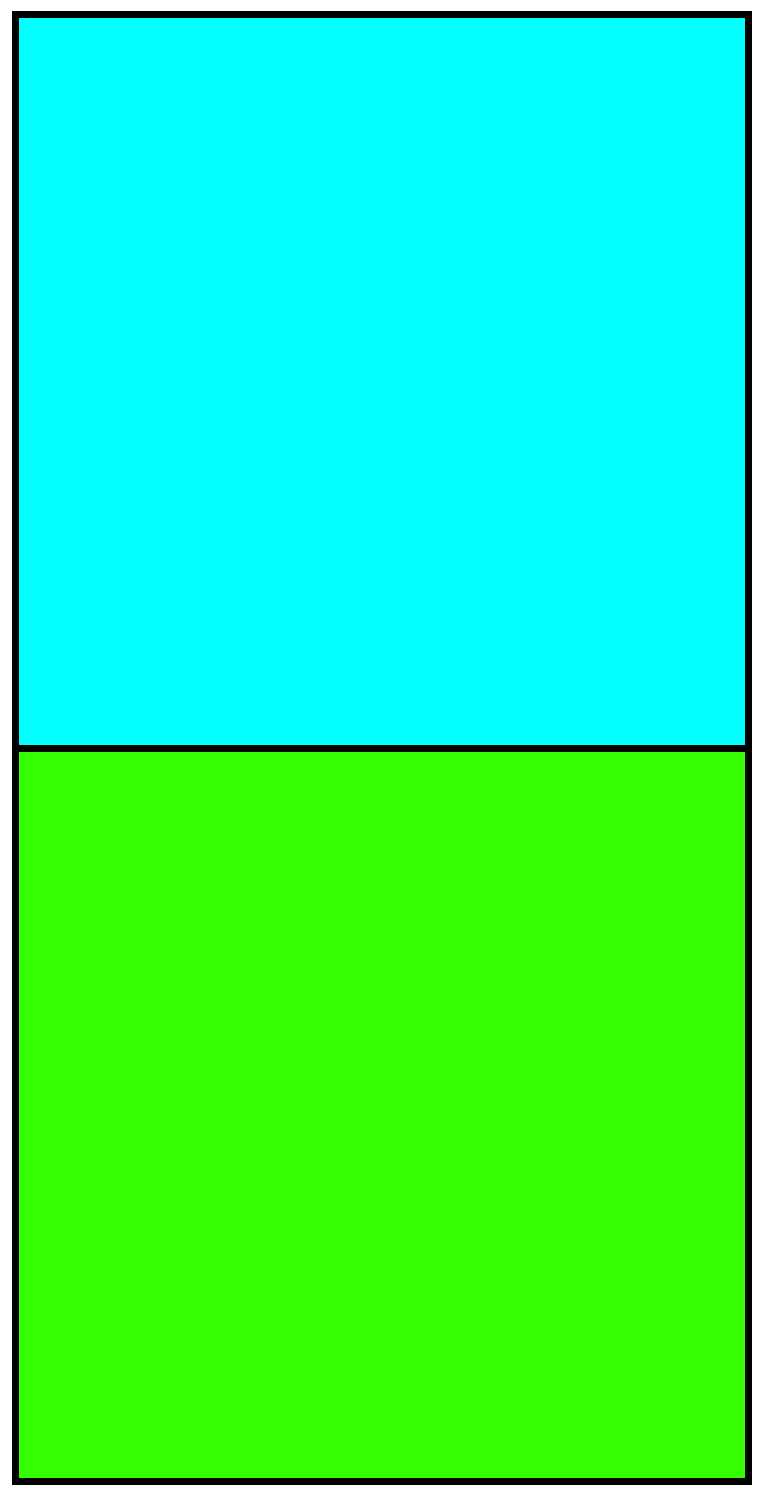}\hfill \includegraphics[width=.2in]{Fib-a}\hfill \\
\caption{The four 1-supertile types.}
\label{onetiles}
\end{figure}

\begin{figure}[ht]
\includegraphics[width=.7in]{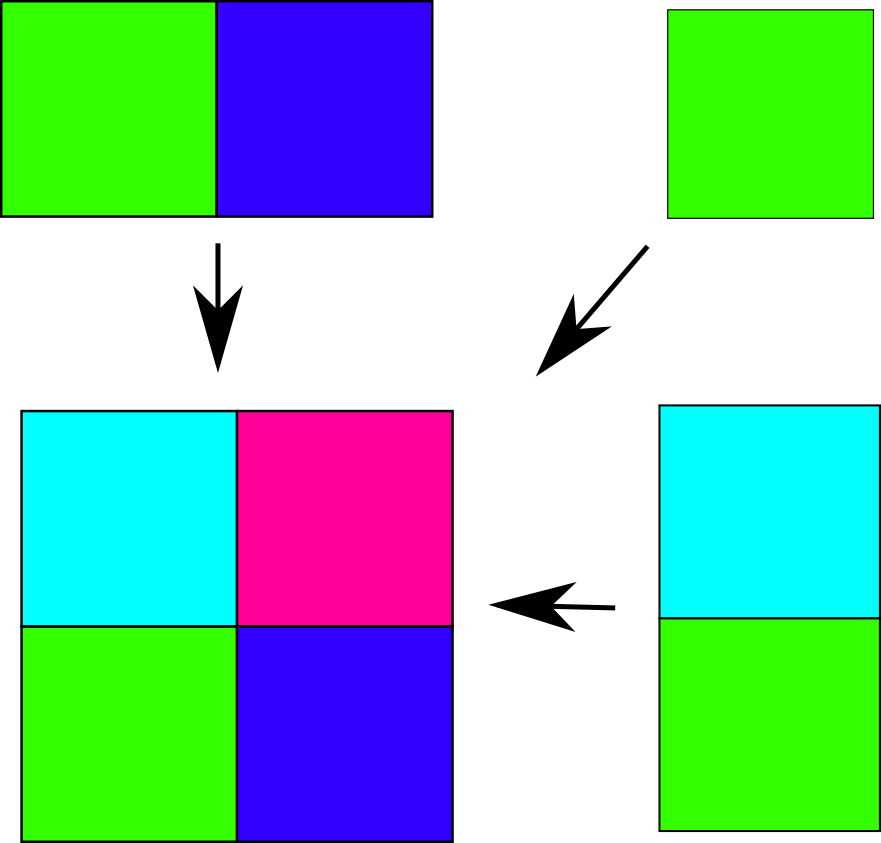} \, \raisebox{.1in}{$=$}  \includegraphics[width=.5in]{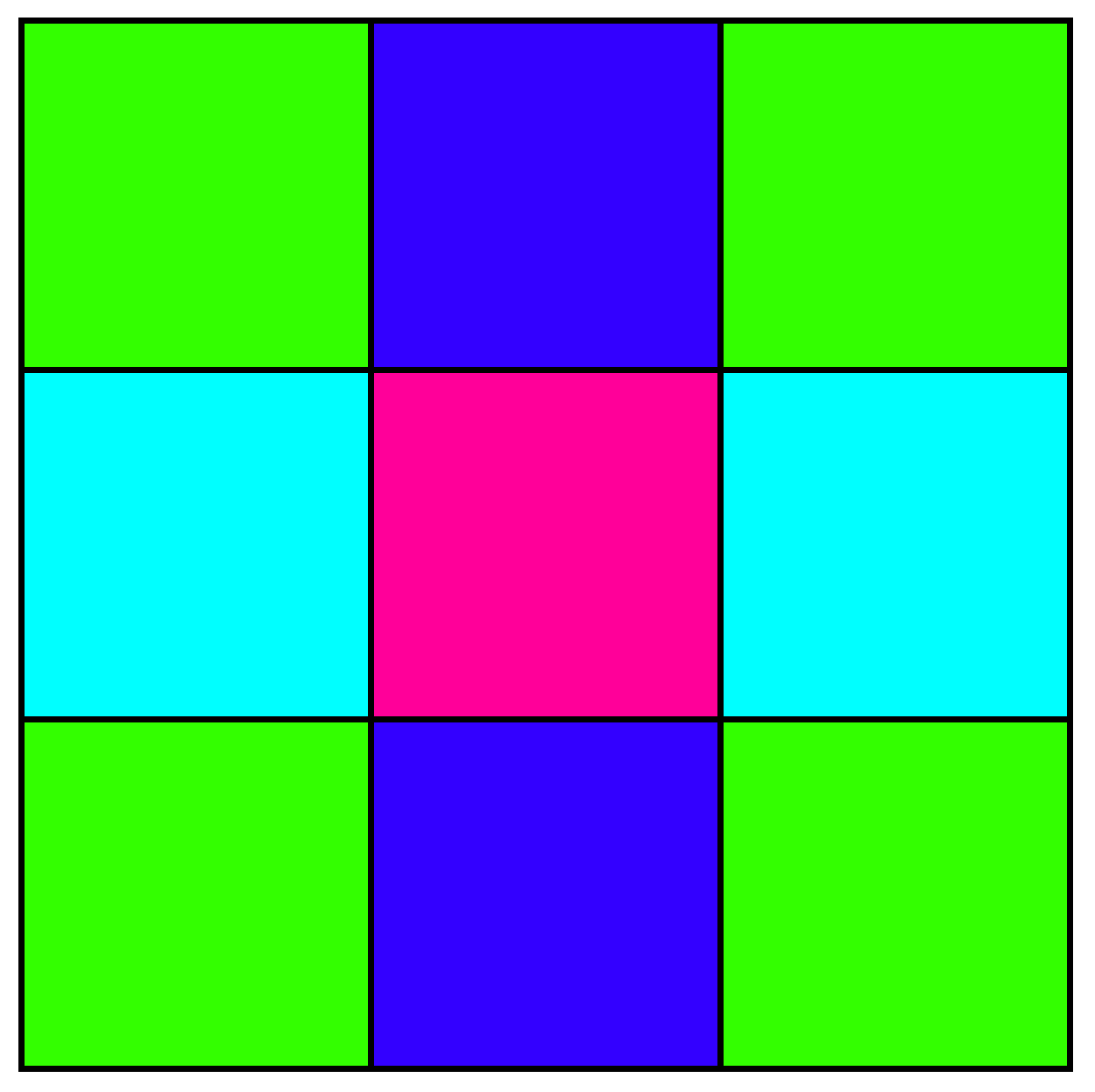}\hfill  %\pause
\includegraphics[width=.7in]{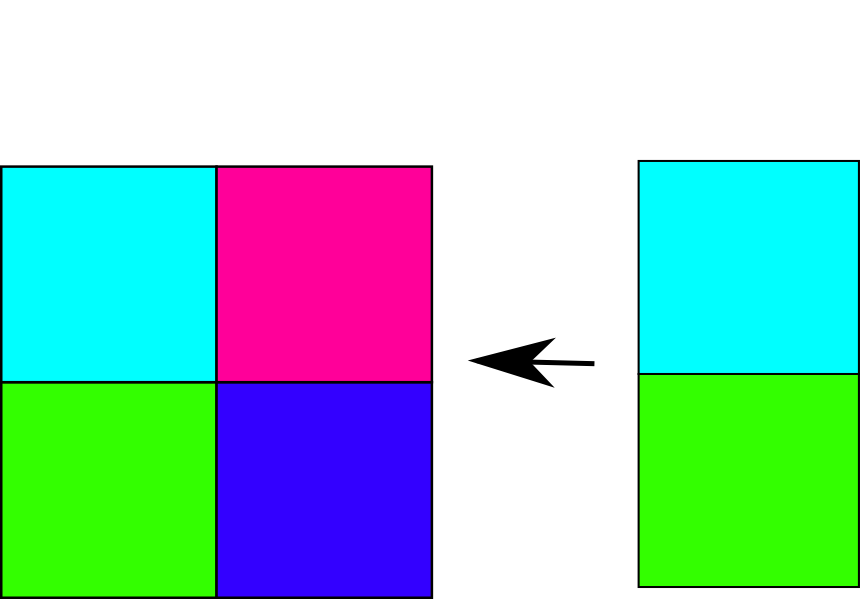} \, \raisebox{.17in}{$=$}  \includegraphics[width=.5in]{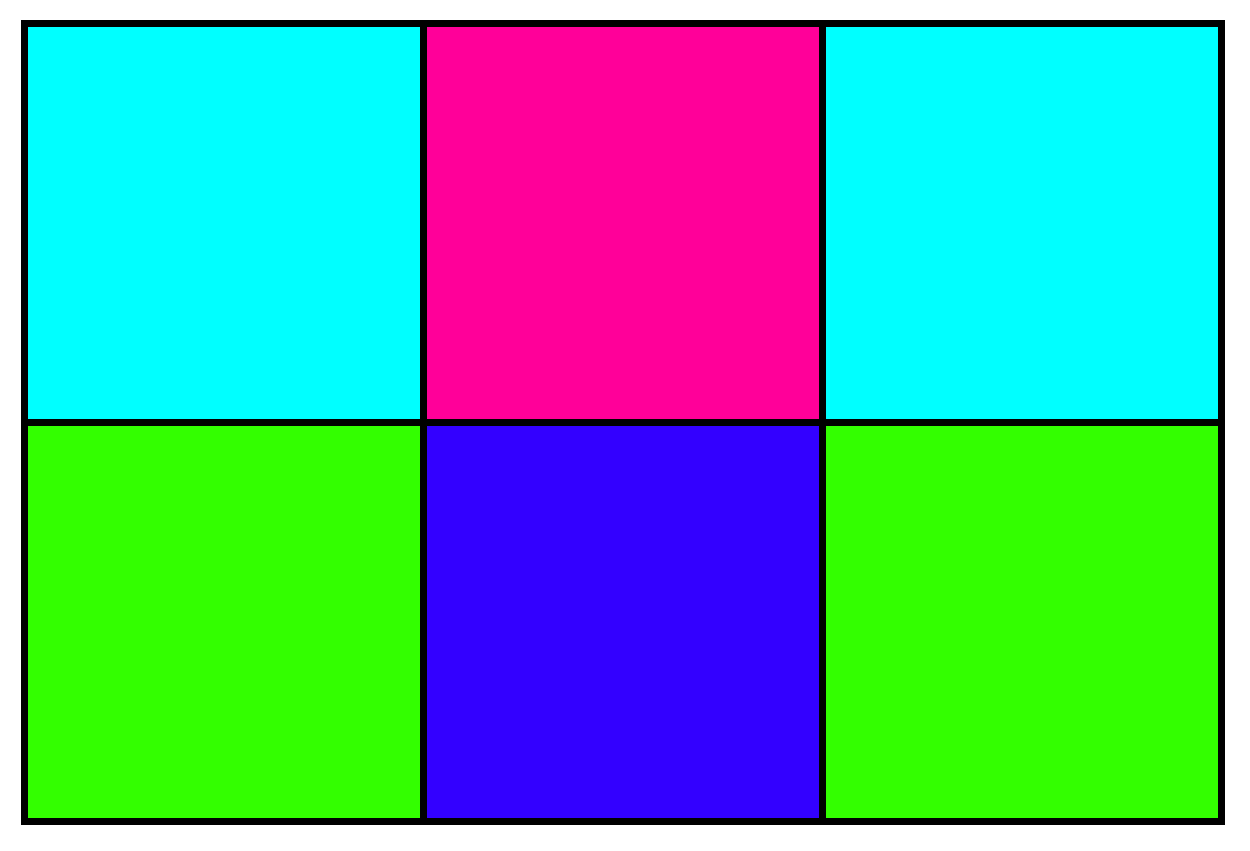}\hfill 
\includegraphics[width=.7in]{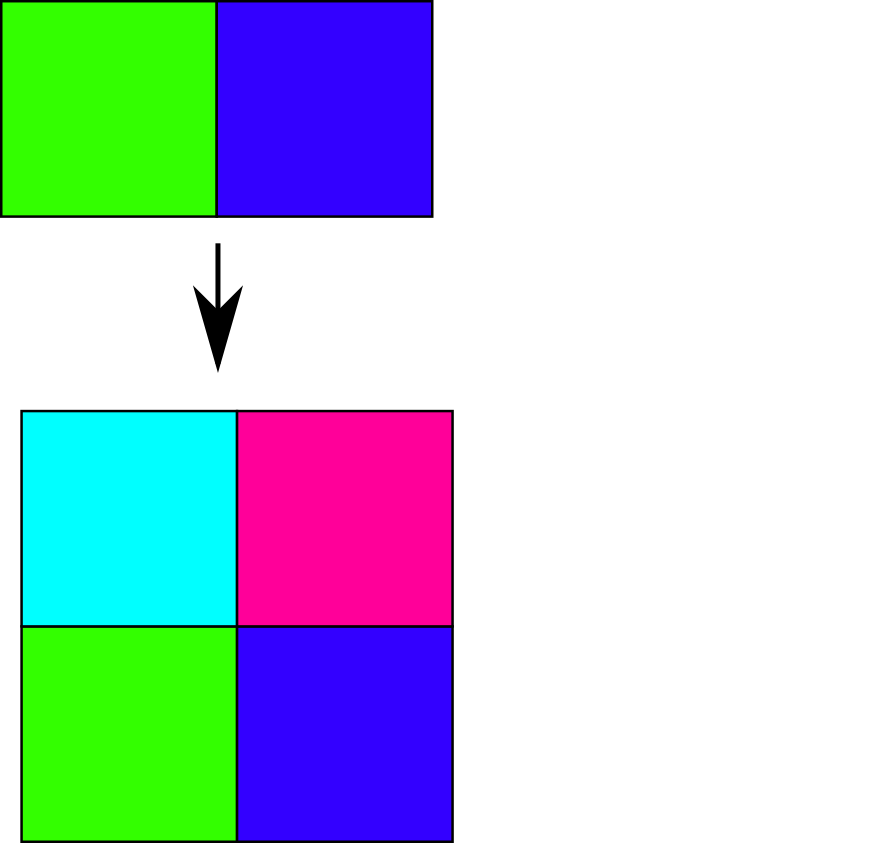} \raisebox{.17in}{$=$} \, \includegraphics[width=.3in]{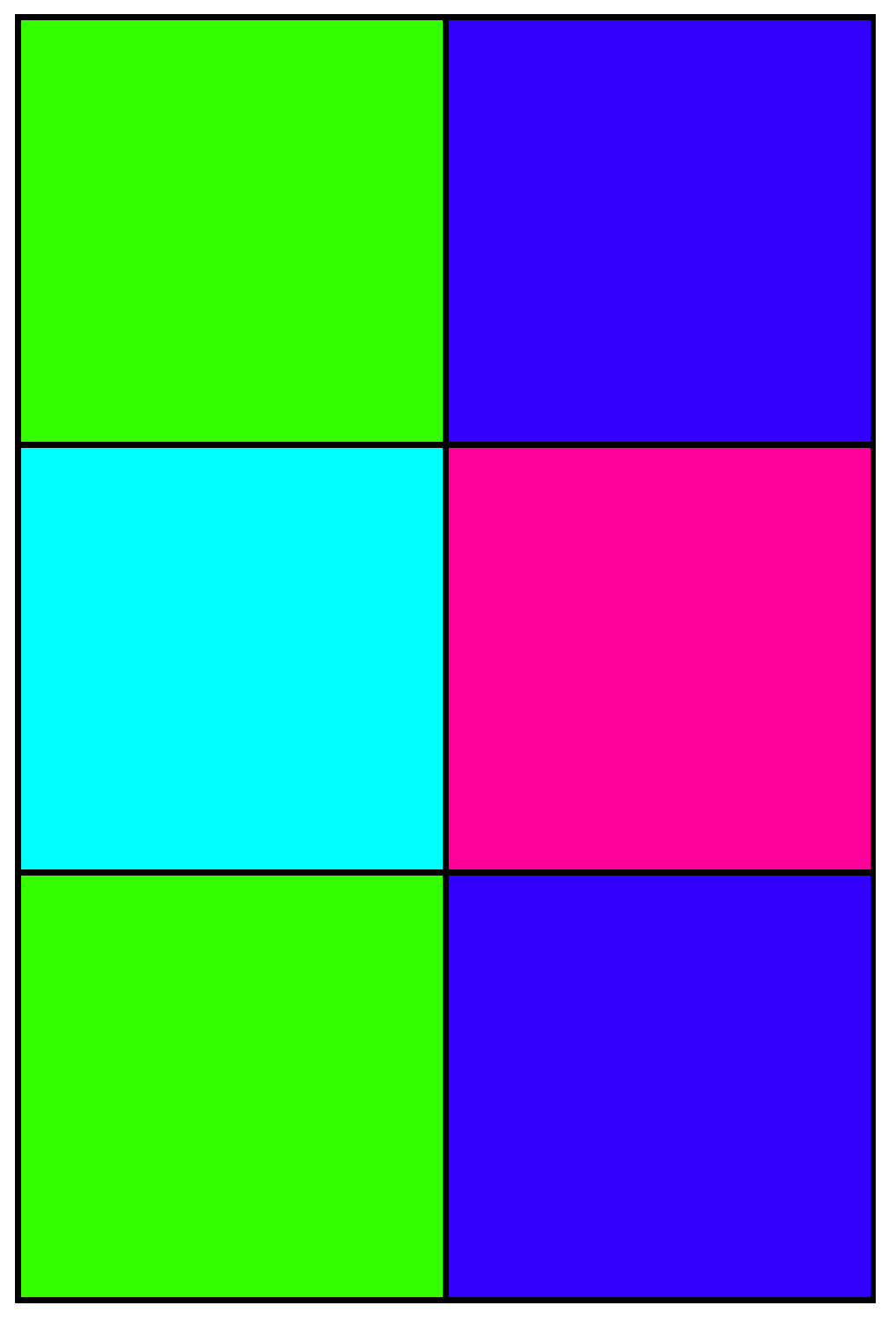}\hfill %\pause 
\caption{How the 1-supertiles fuse to form 2-supertiles.}
\label{twotilefusion}
\end{figure}

\begin{figure}[ht]
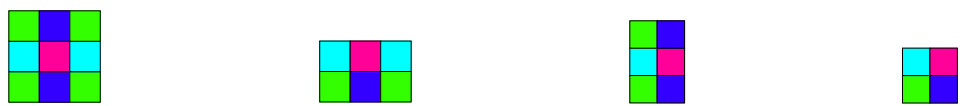

\includegraphics[width=.5in]{fibdp-as2}\hfill \includegraphics[width=.5in]{fibdp-bs2}\hfill \includegraphics[width=.3in]{fibdp-cs2}\hfill \includegraphics[width=.3in]{fibdp-as1}\hfill
\caption{The four 2-supertiles.}
\label{fibtwosupertiles}
\end{figure}

\begin{figure}[ht]
\includegraphics[width=1in]{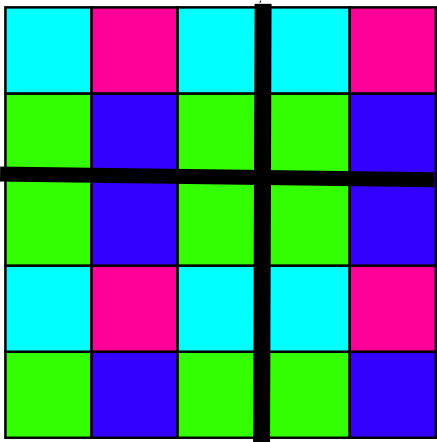}\hfill \includegraphics[width=1in]{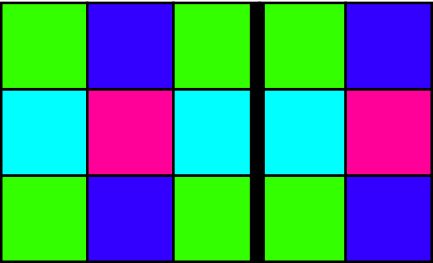}\hfill 
\includegraphics[width=.6in]{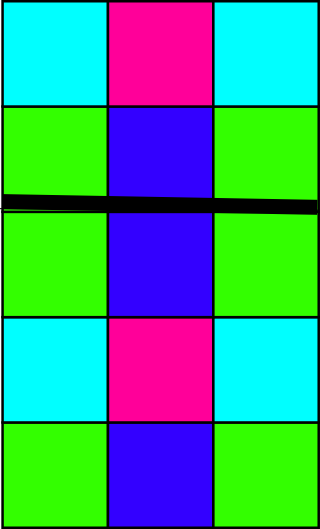}\hfill \includegraphics[width=.6in]{fibdp-as2}\hfill \\
\caption{The four types of 3-supertiles.}
\label{fibthree}
\end{figure}

\begin{figure}[ht]
\includegraphics[width=3in]{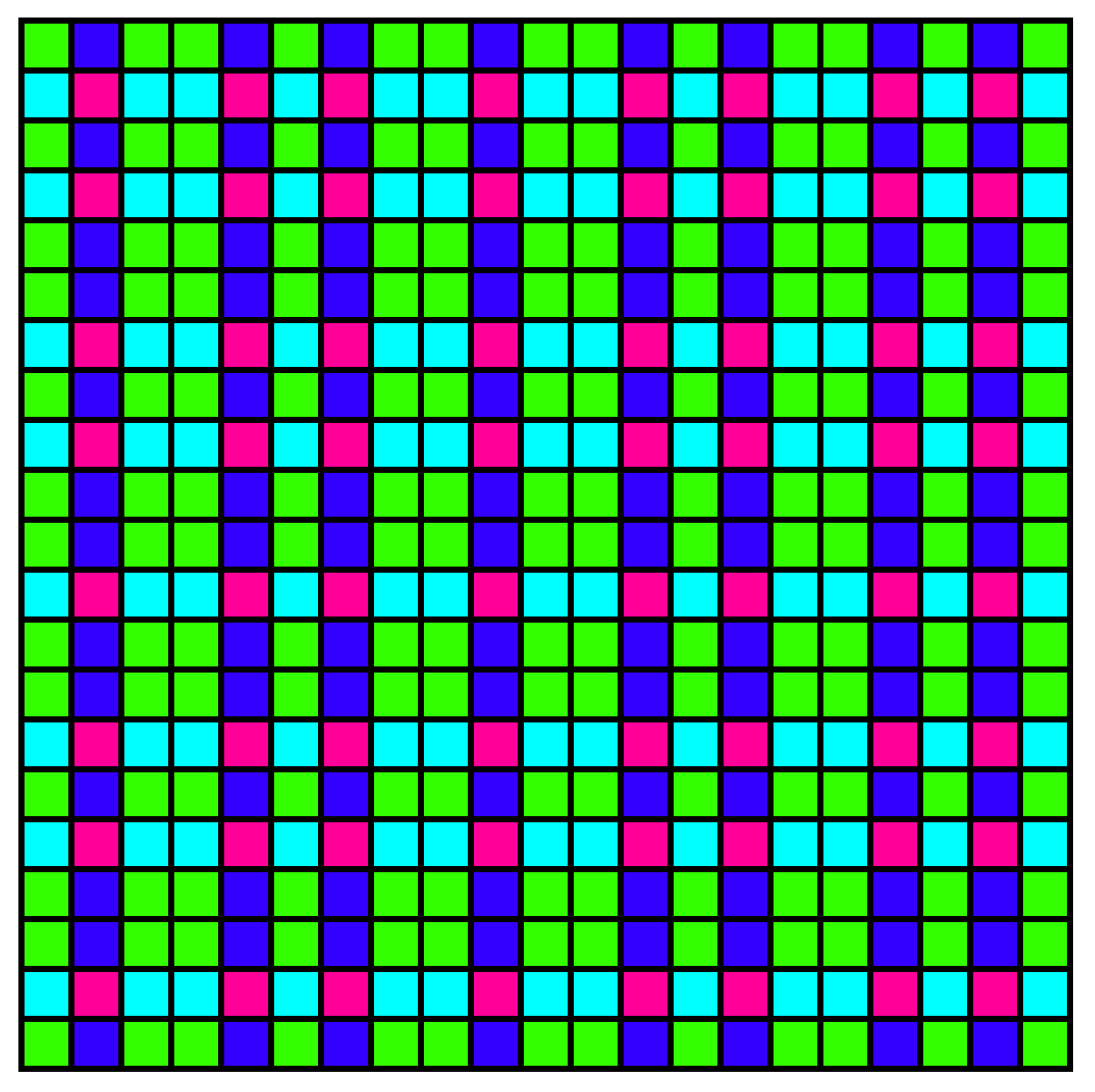}
\caption{A two-dimensional Fibonacci tiling.}
\label{2Dfibpatchlarge}
\end{figure}

Figure \ref{onetiles} shows how prototiles are fused to form four different types of 1-supertiles, again thought of as unit cells.
Figure \ref{twotilefusion} shows our choice of how 1-supertiles fuse together to form three types of 2-supertiles, and in figure \ref{fibtwosupertiles} we show the resulting set of 2-supertiles.  (Note that the fourth 2-supertile is a trivial fusion of a 1-supertile with the empty patch.  This is allowed in the fusion paradigm.)
Figure \ref{fibthree} shows the 3-supertiles with thick lines show the boundary between 2-supertiles for each.
Finally, in figure \ref{2Dfibpatchlarge} we show a large patch of tiles from the fusion rule.  It is possible to show that this tiling is the direct product of two Fibonacci substitution sequences.  The spectrum of the Schr\"odinger operator associated with a closely related tiling space is computed in \cite{RonLif}.

\section{Definitions and further examples}

Before we begin our definitions for fusion tilings we offer references for readers who are interested in going deeper into the motivation of fusion rules.   An excellent resource for tilings in general and their connection with quasicrystals is \cite{Baake-Grimm-book}.
An exhaustive survey of substitution sequences such as the Fibonacci sequence appears in \cite{Fogg}.  The paper \cite{Robinson.ams} serves as a tutorial on substitutions for tilings such as the self-similar Penrose tilings.  
Various forms of tiling substitution rules are compared and contrasted in \cite{Expositiones}.
Fusion rules serve as a physically plausible generalization of these topics and these references provide insight into their mathematical underpinnings.

Now for the definitions.  It is standard to assume that tiles have reasonable shapes such as `topological disks', which are deformed copies of the standard unit disk $\{\vecx \in \R^d \text{ such that } |{\bf x}|\le1\}$\footnote{We restrict our discussion to one and two dimensions but note that everything we are doing has higher-dimensional analogues.}. In one dimension, tiles are closed intervals; in two dimensions they can be polygonal or have fractal edges, but they cannot be disconnected or have holes.  In the event that there are congruent tiles that we wish to distinguish, tiles can carry labels.  We follow the convention that requires that there is some finite number of tile types, congruent copies of which are used to tile the plane without gaps or overlaps.

It is often convenient to think about {\em patches} of tiles:  finite collections of tiles that cover a connected set and overlap only on their boundaries.  The fundamental action of sticking two patches together to form another, larger patch is called `fusion', which we think of  as a geometric generalization of the idea of concatenation:

\begin{dfn} A {\em fusion} of a patch $P_1$ to another patch $P_2$ is a union
of $P_1$ and $P_2$ that is connected and does not contain overlaps.  
\end{dfn}

Notice that the fusion of tiles is always a patch, and that patches can be fused to congruent copies of themselves.  The only rule is that the end result is a finite union of tiles covering a connected set and intersecting only on their boundaries.

To define a fusion rule we build up supertiles, special patches of tiles, in a series of levels:

\begin{itemize}
\item {\em 0-supertiles.}
A finite collection $\ppp_0$ of tiles, often called `prototiles'.
\item {\em 1-supertiles.} A finite collection $\ppp_1$ of patches (fusions) of tiles from $\ppp_0$.
\item {\em 2-supertiles.} A finite collection $\ppp_2$ of patches made by fusing together $1$-supertiles from $\ppp_1$.
\item {\em $n$-supertiles.} For each $n$, $\ppp_n$ is a finite set of patches that are fusions of $(n-1)$-supertiles.

\end{itemize}

\begin{dfn}
Suppose that the number of $n$-supertiles is denoted $j_n$ and write $\ppp_n = \{P_n(i) | 1 \le i \le j_n\}$, where the ordering of supertiles does not matter.
A {\em fusion rule} is the collection of all supertiles: $\rrr = \{\ppp_n \text{ such that } n \in \N\} = \{P_n(i) \text{ such that } n \in \N \text{ and } 1 \le i \le j_n\}$. 

\end{dfn}

A fusion rule defines all possible allowed patches for tilings in a fusion tiling space $\Omega$ as follows. Consider an infinite tiling $\T$ and suppose $\rrr$ is a fusion rule.  We say $\T$ is a {\em fusion tiling with fusion rule $\rrr$} if every patch that appears inside of $\T$ also appears somewhere inside a supertile in $\rrr$.  Think of $\T$ as being constructed using supertiles of unbounded size.

%\pause
\bigskip

\subsection{What kinds of tilings can be seen as fusion tilings?}
\begin{figure}[ht]
\includegraphics[width=2.25in]{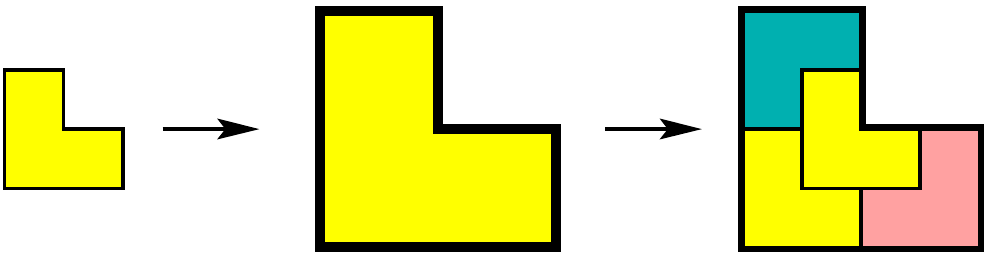}\\
\caption{The chair substitution rule.} \label{chairsubs1.color}
\end{figure}

\begin{figure}[ht]
\includegraphics[width=2in]{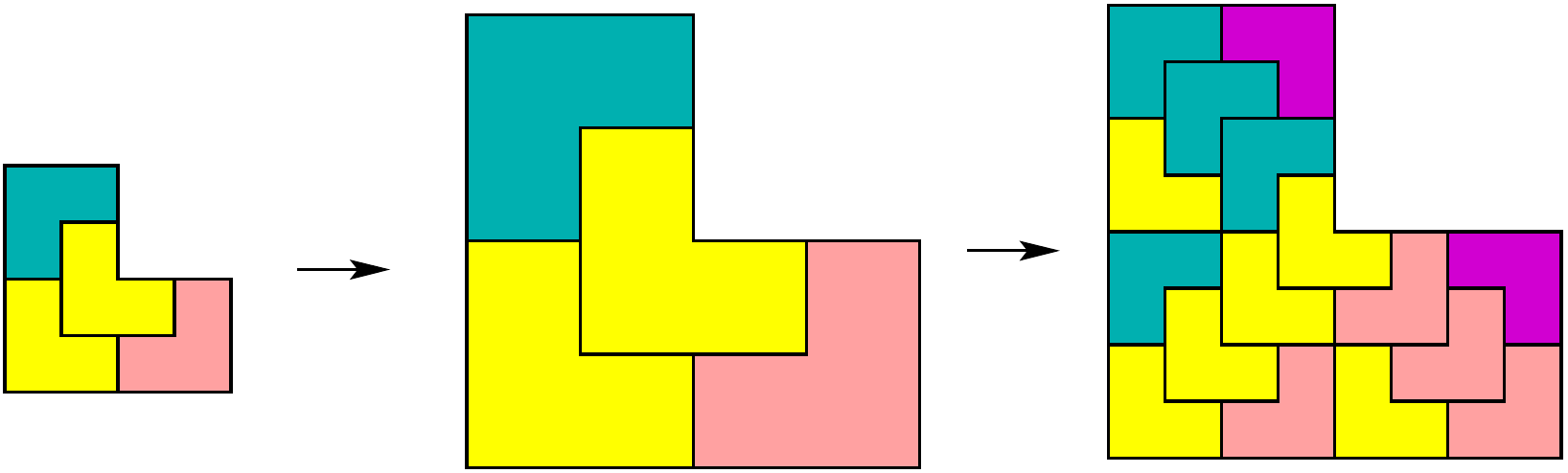}
\caption{A 2-supertile via inflation and subdivision.}\label{chairsupertile.1}
\end{figure}

It is especially important to note that all symbolic substitutions (such as the well-studied Fibonacci substitution) and inflate-and-subdivide rules that generate self-similar tilings (such as the Penrose, octagonal, and pinwheel tilings) can be seen as fusions.  As an example we show how the `chair' inflate-and-subdivide rule of figure \ref{chairsubs1.color} is seen as a fusion.  To produce a 1-supertile, the chair tile is inflated and subdivided.  Figure \ref{chairsupertile.1} shows how a 2-supertile is obtained by inflating and subdividing all of the tiles of a 1-supertile.
The inflate-and-subdivide process can be repeated to obtain higher-order supertiles.

This model can be thought of as a `cellular': tiles are cells that grow until they are the right size to be subdivided. 

\begin{figure}[ht]
\includegraphics[width=1.0in]{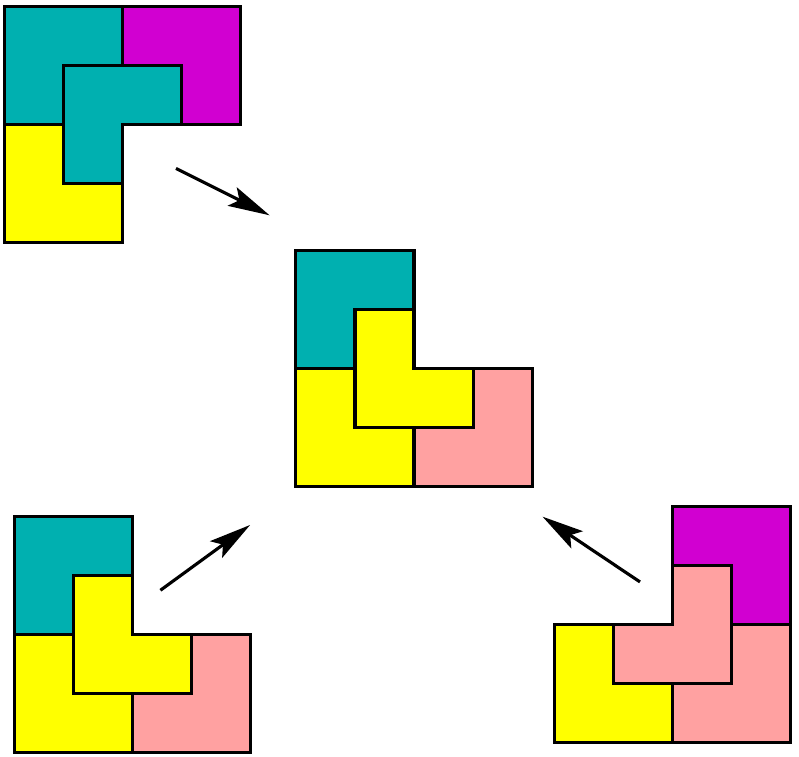}\hspace{.8cm} \raisebox{.5in}{$=$} \hspace{.8cm} \raisebox{.2in}{\includegraphics[width=.625in]{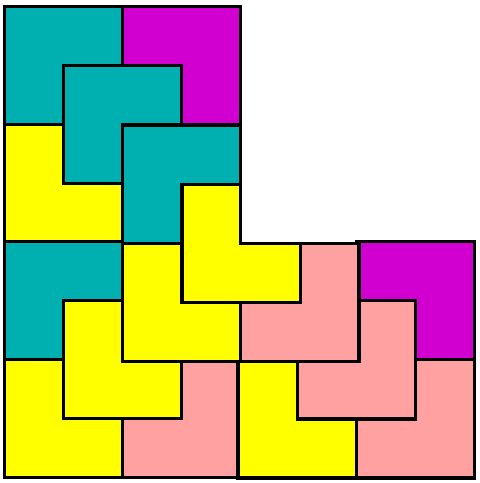}}
\caption{A 2-supertile via fusion.}\label{fusion.chairsupertile}
\end{figure}

By way of contrast, suppose $\ppp_1$ contains the 1-supertile from figure \ref{chairsubs1.color} thought of as a fusion of four prototiles. 
 Figure \ref{fusion.chairsupertile} shows how to make the 2-supertile of figure \ref{chairsupertile.1} as a fusion of  1-supertiles.

The hallmark of inflate-and-subdivide rules is geometric rigidity: each supertile is an expansion of the last by a given factor.
This can also happen for fusion rules, but it is not necessary.

In fact  it is possible to construct any given tiling as a fusion tiling \cite{MeLorenzofusion} unless we impose further restrictions on a fusion rule.  To get interesting results, and especially to get quasicrystalline tilings, it is necessary to require some additional properties.  One is to require that the fusion rule is ``van Hove" in the sense of definition \ref{vanHove} below; this restriction forces $k$-supertiles to grow in size as $k \to \infty$ in a reasonable way and is automatically satisfied by inflate-and-subdivide rules.  Another is ``primitivity", an important idea from substitution systems that is adapted to the fusion case in definition \ref{primitivity}.

Even if a fusion rule is van Hove and/or primitive, it is still possible to think of examples that result in periodic tilings or where the decomposition of a given tiling into its constituent $k$-supertiles is not unique.  For the sake of exposition we do not go into these technical details and refer the reader to \cite{MeLorenzofusion}.

\subsection{Two more examples}
The fusion paradigm is quite flexible and has the following attributes:
\begin{itemize}
\item defects can be accounted for,
\item rules can change from level to level or break down after a certain scale is reached,
\item the number and shape of supertiles can vary from level to level, and
\item randomness can be present.
\end{itemize}

\begin{ex}[A Fibonacci-like example]\label{Fiblike}
In this example we show the flexibility of the fusion paradigm by allowing the number of supertiles to vary from level to level.
Begin  with prototile set $\{\reda,\blueb,\greent\}$ and denote  the $k$-supertiles  by  $A_k, B_k, T_k$.  We have chosen the fusion rule in such a way that supertiles of the $T$ type appear only in levels indexed by $k = 3^n - 1$.  The fusion rules are as follows:
\begin{itemize}
\item $\blueb_k  = \reda_{k-1}$ for all $k$
\item if $k \neq 3^n$ then $ \reda_k  = \reda_{k-1} \blueb_{k-1}$ and if  $k = 3^n, \reda_k  = \greent_{k-1} \blueb_{k-1}$
\item if $k = 3^n-1$ then $\greent_k  =\blueb_{k-1}\reda_{k-1}.$
\end{itemize}

In the following table we compute the $k$-supertiles for $k = 0$ to 4, putting underbraces to emphasize the $(k-1)$-supertiles that compose the $k$-supertiles.

\begin{tabular}{|m{3cm}|m{2.5cm}|m{2cm}|m{1cm}|}\hline%\vspace{.5cm}
 $\reda$ & $\blueb$ & $\greent$&$k=0$ \\\hline %\pause 
%\vspace{.5cm}
 $\greent \blueb$ & $\reda$ & & $k=1$ \\\hline%\pause%\vspace{.5cm}
  $\underbrace{\greent \blueb} \reda$ & $\underbrace{\greent \blueb} $& $\reda\underbrace{ \greent \blueb}$  & $k=2$ \\\hline
  %\pause
 % \vspace{.5cm}
 $\underbrace{\reda\greent\blueb}\underbrace{\greent\blueb}$ & $\underbrace{\greent\blueb\reda}$ &  & $k=3$ \\\hline %\pause
 $\underbrace{\reda\greent\blueb\greent\blueb}\underbrace{\greent\blueb\reda}$&$\underbrace{\reda\greent\blueb\greent\blueb}$&&$k=4$\\ \hline
  \end{tabular}

\end{ex}

\begin{ex}[The $10^n$ example]\label{10^n}
In this example we let the prototile set be $\ppp_0 = \{A,B\}$ and denote the $n$-supertiles by  $\ppp_n =  \{A_n, B_n\}$.
This time we let the number of supertiles at each level remain the same but vary the fusion rule by level.  The 1-supertiles are given by
$A_1 = AAAAAAAAAAB = A^{10}B$ and $B_1 = BBBBBBBBBBA= B^{10}A$; the 2-supertiles by
$A_2 = (A_{1})^{100} B_{1}$ and $B_2 = (B_{1})^{100}A_{1}$, and the $n$-supertiles by
$A_n = (A_{n-1})^{10^n} B_{n-1}$ and $B_n = (B_{n-1})^{10^n}A_{n-1}$.  

Tilings admitted by this substitution will have a much higher frequency of $A$'s than $B$'s in all supertiles of type $A$ but the situation is reversed in all supertiles of type $B$.  However, each supertile type will contain large numbers of both and the space is homogeneous in the sense of theorem \ref{MeLorenzoMinimality}.
\end{ex}

%We will revisit these fusions shortly as they provide nice examples for computation.

\section{Transition matrices}
Computing the number of tiles of various types in any particular $k$-supertile is done using {\em  transition matrices}, a process similar to one used in substitution sequences and tilings.  In that context there is a single matrix $M$ with nonnegative integer entries where the entry $M_{ij}$ is the number of tiles of type $i$ in the substitution of the tile of type $j$.  The number of tiles of type $i$ in the $n$th substitution of $j$ is given by the $(i,j)$ entry of $M^n$.  Since fusion rules can change from level to level, we need a family of transition matrices that tell us the composition of $n$-supertiles inside of $N$ supertiles.  Recall that $j_k$ denotes the number of $k$-supertiles and that we use the notation $\ppp_k = \{P_k(1), P_k(2), ... P_k(j_k)\}$.

\begin{dfn}
For natural numbers $0 \le n < N$, the {\em transition matrix} $M_{n,N}$ counts how many of each type of $n$-supertile make up the $N$-supertiles; its $(i,j)$ entry is given by
$$M_{n,N}(i,j)= \#(P_n(i) \hbox{ in } P_N(j)),$$ where $1 \le i \le j_n$ and $1 \le j \le j_N$. 
\end{dfn}

The notation $\#(P_n(i) \hbox{ in } P_N(j))$ means the number of copies of the supertile $P_n(i)$ in the supertile $P_N(j)$.  There is the possibility that copies of $P_n(i)$ overlap in a nontrivial way, for instance if the fusion rule allows periodic tilings.   Since most examples of interest do not have this problem we will ignore it and refer the reader to \cite{MeLorenzofusion} for the strategy for dealing with such a situation.

The columns of transition matrices can be thought of as population vectors, sorting the population of $n$-supertiles in an $N$-supertile by type.   The $j$th column tells us how $P_N(j)$ is populated by $n$-supertiles.

Whenever $0 \le n < m < N$ it is true that $M_{n,N} = M_{n,m}M_{m,N}$, since the number of 
$n$-supertiles in an $N$-supertile can be computed by the number of $n$-supertiles in the $m$-supertiles and then accounting for how many $m$-supertiles make up that $N$-supertile.  This means that we really only need to compute $M_{n, n+1}$ for all $n$ to have full information on transitions.

\begin{ex*}[\ref{Fiblike}, continued] Recall that the fusion rules in this example are the same as for the Fibonacci substitution except at levels $3^n - 1$ and $3^n$.  This means that the transition matrices are usually two-by-two, and occasionally three-by-two or two-by-three.
If $k \neq 3^n-1$ or $3^n$, then there are only two tile types at both levels and we obtain the standard Fibonacci matrix
$M_{k-1,k} = \begin{bmatrix} 1&1 \\1&0 \end{bmatrix}$.
If $k = 3^n-1$ we have introduced the third supertile type and thus we have three columns, but only two rows since there are only two tile types at level $k-1$.  We compute
$M_{k-1,k} \begin{bmatrix} 1 & 1 & 1\\1&0&1 \end{bmatrix}$.
Finally, if $k=3^n$ we are eliminating the third tile type that was present at level $(k-1)$ and we obtain the $3 \times 2$ matrix
$M_{k-1,k} =\begin{bmatrix}0&1\\1&0\\1&0  \end{bmatrix}$.
\end{ex*}

\begin{ex*}[\ref{10^n}, continued] Recall the fusion rule:
$$A_n = (A_{n-1})^{10^n} B_{n-1}\qquad B_n = (B_{n-1})^{10^n}
 A_{n-1}$$  Thus the transition matrices are all of the form
 $M_{k-1,k} =\begin{bmatrix} 10^k&1 \\1&10^k \end{bmatrix}.$
\end{ex*}

\subsection{Primitivity}
The idea of  primitivity for fusions is adapted from the same idea for substitutions and self-similar tilings, where it implies homogeneity in the sense that every block or patch of any size that appears in one tiling appears in all others.
Usage of the term `primitive' comes from the fact that a matrix $M$ with nonnegative entries is said to be primitive if there is a power $M^N$ of the matrix whose entries are all strictly positive.  If $M$ is the transition matrix of a substitution, primitivity implies that every type of $n$-supertile is contained in every type of $n+N$ supertile. 
For fusions primitivity is still a condition that guarantees that every type of $n$-supertile appears in all sufficiently large supertiles.
%To adapt this idea to fusions we need to require that for all $n$-supertiles there is some large enough $N$ such that every $N$-supertile contains a copy of each type of $n$-supertile.  This is as a property of the family of transition matrices:

\begin{dfn}\label{primitivity}
The fusion rule $\rrr$ is {\em primitive} if for every $n$, there exists an $N(n)$ such that all entries of $M_{n,N'}$ are strictly positive whenever $N' \ge N(n)$.
\end{dfn}

A subtle detail that separates primitivity for fusion from substitution is that the question of how large $N$ must be relative to $n$ changes with $n$, whereas for substitutions a single $N$ works for all levels of supertiles.

Examples \ref{Fiblike} and \ref{10^n} are primitive, the first with $N = n+2$ and the second with $N=n+1$ for all values of $n$.   In the terminology of dynamical systems, primitivity implies {\em minimality}: each tiling can be arbitrarily well approximated by translates of any other tiling.  Thus the translational orbit of any given admitted tiling is dense among all admitted tilings.  We have the following theorem.

\begin{thm}{\cite{MeLorenzofusion}} \label{MeLorenzoMinimality} Let $\rrr$ be a primitive fusion rule and suppose $\T$ and $\T'$ are any two tilings admitted by $\rrr$.  Then every patch of tiles found in $\T$ can be found in $\T'$ and vice versa.
\end{thm}

Thus primitivity imparts a form of homogeneity into the tiling space since all tilings locally look like one another.  However, it does NOT imply consistent patch frequency from tiling to tiling.  For example, patch frequencies in the $10^n$ fusion, which is primitive, depend on where you are looking and/or in which tiling.

\section{Frequency computations}

We begin this section by defining what it means for a fusion rule to be van Hove, a concept that is widely useful in the study of fusion tilings.  Supertiles in such a fusion grow large in a `round' way and do not become arbitrarily long and skinny.  One way to ensure this is to require that their boundaries, when padded by some small amount, become trivial in size relative to their interiors.  Let us make this precise.

 Consider a sequence $\{A_k\}$ of subsets of $\R^2$, and for any $r>0$ denote by $\partial A_k^r$ the set of all points in $\R^2$ that are within $r$ of the boundary of $A_k$.  We say that  $\{A_k\}$ is a {\em van Hove} sequence if $\lim_{k \to \infty} \frac{Vol(\partial A_k^r) }{Vol(A_k)} = 0$.

%To ensure that the supertiles in a given fusion rule grow in this somewhat controlled fashion we consider all possible sequences
%constructed by making a choice of $1$-supertile $P_1(i_1)$, then a choice of $2$-supertile $P_2(i_2)$, and so on.
\begin{dfn}\label{vanHove}
A fusion rule $\rrr$ is said to be {\em van Hove} if any sequence of supertiles $\{P_k(i_k)\}_{k = 1}^\infty$ forms a van Hove sequence in $\R^2$.
\end{dfn}
\noindent Note that the sizes of all $k$-supertiles grows without bound as $k$ does whenever the fusion rule is van Hove.

Now we turn our attention to computing relative frequencies of tiles and patches.  The definition for general spaces of tilings is as follows.  Let $\T$ be a tiling in the tiling space and $P$ be a patch of tiles.  To get the frequency of $P$ in $\T$, count the number of $P$ in $\T$ in larger and larger balls around the origin, normalizing by volume:
$$freq(P) = \lim_{n \to \infty} \frac{\#(P \text{ in } \T \cap B_n(0))}{Vol(B_n(0))}$$
The ergodic theorem guarantees that these frequencies will exist for almost every $\T$ in a tiling space.  Letting $P$ range through all prototiles yields a frequency vector for the prototiles.  Different choices of $\T$ can result in different frequency vectors, in which case there is more than one translation-invariant probability measure\footnote{These are the standard measures in the mathematical analysis of tiling spaces and can be described in terms of patch frequencies.  We refer to such a measure as ``a frequency measure".}  on the tiling space.

Under the right conditions, we can use the structure provided by fusion to our advantage  when computing frequencies.
Suppose that  $A$ is a prototile and let $P_k(i_k)$ represent any choice of $k$-supertile.  A reasonable way to estimate the frequency of $A$ is to use a large value of $k$ and compute
$$freq(A) \approx \frac{\#(A \text{ in } P_k(i_k))}{Vol(P_k(i_k))}.$$ 
That is, we count up how many times $A$ appears in the $k$-supertile and divide by the size of that supertile.
Roughly speaking, the frequency is well-defined if we can choose a sequence of supertiles for which the limit exists as 
$k \to \infty$.  

For a fixed $k$-supertile $P_k(j)$,  the column $M_{0,k}(*,j)$ gives its breakdown into prototiles.  Dividing the column by the volume of $P_k(j)$ we obtain the relative frequency vector for prototiles as seen in  $P_k(j)$.  We call such a vector {\em volume-normalized}; any actual frequency vector for the prototiles will lie in the span of the volume-normalized columns of $M_{0,k}$.  Similarly we can find frequency vectors for the $n$-supertiles, and Theorem \ref{MeLorenzoMeasures} will show that this is the trick to getting patch frequencies for fusion rules.  

We wish to define a sequence $\rho = \{\rho_n\}$, where each $\rho_n \in \R^{j_n}$ represents a relative frequency vector for $n$-supertiles.  
 In order for  $\rho_n(i) $ to equal the frequency
 of the supertile $P_n(i)$ we require 
 \begin{itemize}
\item  {\em volume-normalization:}   $\sum_{i = 1}^{j_n} \rho_n(i)
Vol(P_n(i)) = 1$ for all $n$.  This ensures that the frequency measure is a probability measure.
\item  {\em transition-consistency:}
$\rho_n = M_{n,N} \rho_N$ whenever $n < N$.  
That is, the frequency of an $n$-supertile is consistent with the frequencies of the $N$-supertiles it appears inside.
\end{itemize}

\begin{thm}{\cite{MeLorenzofusion}\label{MeLorenzoMeasures}
Let $\rrr$ be a recognizable\footnote{This property means that the fusion rule is invertible in a certain sense.}, primitive, van Hove fusion rule.  
There
is a one-to-one correspondence between the set of all frequency measures $\mu$ on the fusion tiling space and the set of all volume-normalized and transition-consistent sequences $\{\rho_n\}$ with the correspondence 
that, for all patches $P$, 
\begin{equation} \label{measure.from.frequency}
freq_\mu(P) = \lim_{n \to \infty} \sum_{i = 1}^{j_n} \#\left(P \text{
  in } P_n(i)\right) \rho_n(i)
\end{equation}
}
\end{thm}

In theory, formula (\ref{measure.from.frequency}) allows us to approximate the frequency of an arbitrary patch $P$.  Take a large value of $n$ and count up how many times $P$ appears in each $n$-supertile.  Multiplying those counts by the frequency of the respective $n$-supertiles and adding up the result gives a good approximation of the frequency of $P$.

In the case of substitution sequence and self-similar tilings, primitivity allows the use of the Perron-Frobenius theorem to show that there is only one possible sequence of supertile frequencies, and thus there is only one frequency measure.  This means that all such systems are ``uniquely ergodic" once they are primitive.  

The Perron-Frobenius theorem does not apply to fusions unless $M_{k-1,k}$ is always equal to the same fixed, primitive matrix. 
Examples such as the $10^n$ example admit more than one distinct sequence of supertile frequencies and thus have more than one possible frequency measure.  It is possible to construct fusion rules that have multiple measures, each of which have different spectral types, while preserving primitivity and hence the homogeneity of Theorem \ref{MeLorenzoMinimality}.

\section{Conclusion}
This article has touched mainly on the construction and a few basic properties of fusion rules.  More is known, and much more is unknown about them at the time of this writing.  We include a small sampling of topics and references for the interested reader.

The diffraction of substitution and self-similar tilings has been subject to intense research since the discovery of quasicrystals, and a nice survey of the state of the art appears in \cite{BaakeGrimm2012}.  The study of the diffraction of fusion tilings hasn't really begun, but there are results about the closely related ``dynamical spectrum".  The dynamical spectrum is defined to be the set of eigenvalues of the unitary operator induced by translation on the square-integrable functions of the tiling space; this spectrum is known to contain the diffraction spectrum.  Theorem \ref{MeLorenzoMeasures} tells us that for recognizable, primitive, van Hove fusions, finding frequency measures boils down the linear algebra governing the transition matrices. However, the {\em spectral type} of those measures doesn't, and further study must be done to determine when a fusion rule is purely diffractive or has some continuous spectrum in the background.    In \cite{MeLorenzofusion} there are theorems that govern the presence or absence of some spectra along with an interesting example, the ``scrambled Fibonacci", for which the diffraction spectrum is invisible to those theorems.  

Fusion rules provide an easy framework for constructing examples that have certain properties.  We have seen that some properties of substitutions carry over to fusions and others don't.   When the properties don't carry over new fusion rules arise as counterexamples that illuminate what is possible.  For instance, strong mixing and entropy are possible for fusions but not for substitutions, however this can only happen when the transition matrices are unbounded.  Minimal systems can fail to be uniquely ergodic, as the $10^n$ example shows.  Fusion rules can generate tilings with infinite local complexity, a situation studied in \cite{MeLorenzoILCfusion}.  Ever more exotic examples can be constructed that can still be understood with existing tools.

A number of topological and operator-theoretic techniques used to study substitution tilings can be adapted for use in studying fusion tilings.  Importantly, fusion tiling spaces can often be seen as inverse limits and therefore carry the structure of a $C^*$-algebra.  This has implications for gap-labelling theorems and the study of Schr\"odinger operators; many results from the survey chapter \cite{Damaniketal} apply directly to fusion rules.  Moreover it is possible to study the cohomology of fusion tilings, which is addressed in \cite{MeLorenzofusion,MeLorenzoILCfusion} and surveyed in \cite{Lorenzobook}.

The main advantages, mathematically, to the fusion paradigm is that it gives a unified framework for lots of hierarchical structures: substitution and self-similar systems; systems defined by Bratteli diagrams; S-adic systems; cut-and-stack transformations; and so on.  From a physical standpoint fusion seems to provide a more plausible model for the growth of quasicrystals than inflate-and-subdivide and projection schemes.  Fusion rules are flexible enough to include defects, respect issues of scale, and allow some randomness,  while having enough structure to guarantee our ability to study them and to generate new mathematical quasicrystals.

\end{document}